\documentclass[sigconf]{acmart}

\usepackage{booktabs} 
\usepackage{xspace}
\usepackage{amssymb}
\usepackage{amsmath}
\usepackage{makecell}
\usepackage{booktabs} 
\usepackage{todonotes}
\usepackage{customizedmacros}
\usepackage{workinprogressmacros}  
\usepackage{linkeddatamacros}
\usepackage{graphicx}
\usepackage{epstopdf}
\usepackage{url}
\usepackage{algorithm}
\usepackage[noend]{algpseudocode}
\usepackage{soul}

\usepackage{caption}
\usepackage{subcaption}
\usepackage{multicol}
\setcopyright{none}    

\begin{document}
\title{The importance of being dissimilar in Recommendation}

\author{Vito Walter Anelli$^\star$, Joseph Trotta$^\star$, Tommaso Di Noia$^\star$, \\Eugenio Di Sciascio$^\star$, Azzurra Ragone$^\bullet$}
\affiliation{%
	$^\star$Polytechnic University of Bari\\
	Bari - Italy\\
	{firstname.lastname}@poliba.it\\
	$^\bullet$Independent Researcher\\
	azzurra.ragone@gmail.com
}


\begin{abstract}
Similarity measures play a fundamental role in memory-based nearest neighbors approaches. They recommend items to a user based on the similarity of either items or users in a neighborhood. In this paper we argue that, although it keeps a leading importance in computing recommendations, similarity between users or items should be paired with a value of dissimilarity (computed not just as the complement of the similarity one). We formally modeled and injected this notion in some of the most used similarity measures and evaluated our approach showing its effectiveness in terms of accuracy results.
\end{abstract}

%
%
%


\maketitle

\section{Introduction}
Neighborhood-based approaches have been the first family of algorithms developed for collaborative filtering recommender systems. They identify similar users or items and  they provide users a list of items they could be interested in by exploiting the degree of similarity. Though their have been around for many years, it has been shown that neighborhood-based approaches may perform better than latent model based methods to solve the \TopN recommendation problem \cite{DBLP:conf/iir/Aiolli13,DBLP:journals/tois/DeshpandeK04,DBLP:conf/kdd/KabburNK13,DBLP:conf/icdm/NingK11}.
In the \TopN recommendation task, the focus is on providing an accurate ranked list rather than minimizing the rating prediction error. 
Among neighborhood-based methods the best-known are user-kNN, item-kNN and Sparse LInear Methods (SLIM) \cite{DBLP:conf/icdm/NingK11}. User-based and item-based schemes have proven to be effective in different settings despite they use the same logic behind the scenes. In details item-kNN and SLIM (which uses an item-based scheme) have shown to outperform user-kNN to solve the \TopN recommendation problem \cite{DBLP:conf/recsys/Christakopoulou16} and several algorithms have been proposed in the literature to enhance neighbors models like GLSLIM \cite{DBLP:conf/recsys/Christakopoulou16} and Weighted kNN-GRU4REC \cite{DBLP:conf/recsys/JannachL17}, taking advantage of personalized models and recurrent neural networks.  Moreover, we also have approaches focusing on the injection of time in neighborhood models \cite{DBLP:conf/recsys/BelloginS17} and modeling similarities by directly optimizing the pair-wise preferences error \cite{DBLP:conf/uai/RendleFGS09}. All in all, under the hood what makes neighborhood-based methods work is a similarity measure. 

Several similarity measures have been proposed and used extensively such as Jaccard \cite{dong2011efficient}\cite{qamar2008similarity} and 
Tanimoto \cite{DBLP:journals/eswa/SilvaCPR16} coefficients, Cosine Vector similarity \cite{DBLP:journals/cacm/BalabanovicS97, DBLP:journals/umuai/BillsusP00, abiteboul1995news},  Pearson Correlation \cite{DBLP:journals/ir/HerlockerKR02},  Constrained Pearson correlation \cite{DBLP:conf/chi/ShardanandM95},  Adjusted Cosine similarity \cite{DBLP:conf/www/SarwarKKR01}, Mean Squared Difference similarity \cite{DBLP:conf/chi/ShardanandM95}, Spearman Rank Correlation \cite{kendall1990rank}, Frequency-Weighted Pearson Correlation \cite{DBLP:conf/uai/BreeseHK98}, Target item weighted Pearson Correlation \cite{DBLP:series/sci/BaltrunasR09}. In the vast majority of cases,  all these similarity measures are based on two assumptions: 
\begin{enumerate}
	\item the correlation between $i$ and $j$ is the same correlation between $j$ and $i$ (symmetry of similarity);
	\item the correlation between two entities captures only how much they are similar with each other without taking into account their degree of dissimilarity.
\end{enumerate}
To the best of our knowledge, a few works have been proposed in the past years related to asymmetric similarity, and they are mainly designed for a user-based scheme.  An Asymmetric User Similarity has been proposed in \cite{millan2007collaborative} where the authors underline that a similarity measure should distinguish between a user with a rich profile and a cold user. Thus, given two users $u$ and $v$, they slightly modify the Jaccard index in order to consider exclusively the number of ratings of the current user (instead of the overall number of both users). In HYBRTyco \cite{katarya2016effectivecollaborative} the similarity proposed by Millan \cite{millan2007collaborative} is combined with the S\o rensen index \cite{Sorensen-1948-BK}. HYBRTyco \cite{katarya2016effectivecollaborative} is an hybrid recommender system which combines matrix factorization with an asymmetric similarity model to realize a typicality-based collaborative filtering recommender system. The same approach is exploited in another asymmetric user similarity model \cite{pirasteh2015exploiting} to feed a user-user similarity matrix that is then completed using a matrix factorization algorithm. Additionally, both of them provide an extension for the latter similarity measure based on explicit numerical feedbacks (ratings). 
Despite previous works are focused on the user-based scheme, we already underlined that item-kNN shows very good performance in \TopN recommendation task. Moreover, when the number of users exceeds the number of items, as in most of the cases, item-based recommendation approaches require much less memory and time to compute the similarity weights than user-based ones, making them more scalable. Due to these reasons, both approaches have been considered in this work.

In this work we investigate the effect on recommendation accuracy when we go beyond the above two assumptions and define (and include) the concepts of dissimilarity and asymmetry in similarity measures. In our proposal, we  start from a probabilistic interpretation of similarity to define symmetric and asymmetric dissimilarities. The dissimilarity measures are then combined with traditional similarity values using additive and multiplicative strategies. The experimental evaluation shows that our approach outperforms the non-dissimilarity-aware counterparts improving accuracy of results or diversity or both.

The rest of the paper is organized as follows: 
Section \ref{sec:approach} presents the motivation behind our work and the proposed approach. Section \ref{sec:experiments} presents the evaluation protocol, metrics, datasets and performance of the method. Finally, in Section \ref{sec:conclusion} concluding considerations are provided. 

\section{Proposed approach}\label{sec:approach}
\subsection{Motivation}\label{sec:motivation}
A symmetric similarity may not be sufficient to capture subtle interactions between items. Our assumption is that representing the similarity through traditional measures can lead to imperfect results as important information might not be considered . 
Let us consider some examples in a item-kNN scenario. Suppose we are dealing with a dataset containing ratings data from the book domain on the following books: 

\begin{center}
\begin{tabular}{l l l r}
	\hline
	Title & Short name &  Author &  \# Votes \\
	\hline
	A Game of Thrones & GoT &  G. R. R. Martin &  100\\ 
	A Dance with Dragons &  DwD & G. R. R. Martin  & 10\\
	Shroud of Eternity & SoE &  T. Goodkind &  120\\
	\hline
\end{tabular}
\label{tbl:example_books}
\end{center}

\begin{figure}[ht]
	\centering
	\begin{subfigure}[t]{0.125\textwidth} 
		\includegraphics[width=\textwidth]{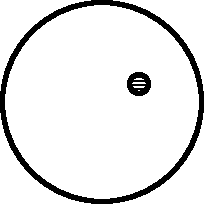}
		\caption{$U_{DwD}$ is entirely contained in $U_{GoT}$} 
	\end{subfigure}
	\hspace{2em} 
	\begin{subfigure}[t]{0.225\textwidth} 
		\includegraphics[width=\textwidth]{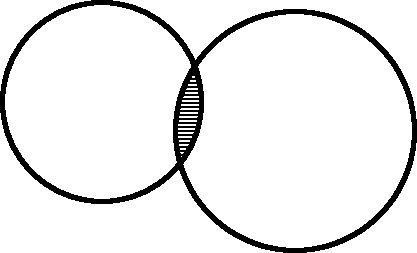}
		\caption{$U_{GoT}$, is partially overlapped with $U_{SoE}$} 
	\end{subfigure}
	\caption{Representation of the motivation example} 
	\label{fig:book_example_sets}
	\vspace{-0.5cm}
\end{figure}

In details both "A Game of Thrones" and "A Dance with Dragons" belong to the same saga "A song of ice and fire" and they are, respectively, the first and the fifth volume. "Shroud of Eternity" is the second volume of Nicci Chronicles's saga. We may assume that all the users who rated DwD also rated GoT, i.e, $U_{DwD} \subseteq U_{GoT}$. Analogously, we may assume that a number of readers of SoE also voted GoT, $U_{SoE} \cap U_{GoT} \neq \emptyset$. Suppose now that we have $U_{SoE} \cap U_{GoT} = 20$ and $U_{DwD} \cap U_{GoT} = 10$. If we compute the Jaccard similarity between the pairs  SoE, GoT and DwD, GoT we have
\begin{eqnarray*}
JS(DwD, GoT) & = & \frac{|U_{DwD} \cap U_{GoT}|}{|U_{DwD} \cup U_{GoT}|} = 0.1\\
JS(SoE, GoT) & = & \frac{|U_{SoE} \cap U_{GoT}|}{|U_{SoE} \cup U_{GoT}|} = 0.1\\
\end{eqnarray*}
In our opinion, a lot of relevant information has been lost in this simple example. The scenario is shown graphically in Figure \ref{fig:book_example_sets}. 

It is clear that $U_{DwD}$ is a proper subset of $U_{GoT}$, on the contrary there are a lot of users in $U_{SoE}$ that have not experienced GoT. This information is mostly lost after the similarity measurement even though a piece of this information is retained in the denominator of the Jaccard coefficient through the overall value of $U_{GoT} \cup U_{DwD}$ and $U_{GoT} \cup U_{SoE}$. Let us focus on a more formal description of the scenario to explicit the reason why we do not consider this remaining information sufficient.

Many similarity measures (like Jaccard in the previous example) mainly rely on a value that denotes the similarity between two items normalized by their overall weight. This can be represented as a probability. Let $\mathcal{I}(u) = \{\langle i,r_{ui}\rangle \mid u \text{ rated } i \text{ with } r_{ui}\}$ be the user profile containing the pairs item-rating and $\mathcal{U}(i) =  \{\langle u,r_{ui}\rangle \mid \langle i,r_{ui}\rangle \in \mathcal{I}(u)\}$ be the set of users that experienced $i$. If we consider Jaccard similarity we see it represents the following probability:
\begin{equation}\label{eqn:sym_prob}
p^{JS}(i,j) =	p(\langle {u},r_{{u}i} \rangle \in \mathcal{U}(i) \wedge \langle {u},r_{{u}j} \rangle \in \mathcal{U}(j) )
\end{equation} 
which can be read as the probability that $u$ experienced both $i$ and $j$. Analogously, we may define a dissimilarity measure, given by two different probabilities:
\begin{itemize}
	\item the probability that a generic user $u$ experienced the item $i$ but never experienced the item $j$
	\item the probability that a generic user $u$ experienced the item $j$ but never experienced the item $i$
\end{itemize}
This consideration is certainly important but these two different probabilities are not equally important.

In a classical memory-based Item-kNN, $sim(i,j)$ is used to compare $i$ to different $j$ to find out which $j$s are the most similar to $i$. 
In practical terms, we are interested in how much $j$ is similar to $i$. 
If we focus on Figure \ref{fig:book_example_sets}a, we realize that DwD is more similar to GoT than the opposite situation. The reason behind this behavior  is not directly related to the size of the involved sets but it depends on the probability that a user who experienced GoT did not experience DwD. 

Though some interesting asymmetric similarities have been proposed in the last years, to our knowledge, no one focused on this probability that represents a negative asymmetric dissimilarity. 

\subsection{Metrics}
In this work we propose a  general asymmetric similarity model in which items $i,j$ similarities are computed by taking into account the probability that users that experienced $j$ never experienced $i$. 
The idea behind our work is preserving the core meaning of a specific similarity, applying a corrective factor encoding the dissimilarity we mentioned before.

We introduced this correction into two binary symmetric similarities: Jaccard and S\o rensen index, and in the asymmetric variant to Jaccard coefficient proposed in the literature \cite{millan2007collaborative}. We tested this correction both as an additive and as as a multiplicative factor. 


For the sake of completeness, we reintroduce \textit{Jaccard coefficient similarity (JS)} that, for a memory-based item-kNN model can be expressed as:
\begin{equation}\label{eqn:js}
JS(i,j)=\frac{|\mathcal{U}(i) \cap \mathcal{U}(j)|}{|\mathcal{U}(i) \cup \mathcal{U}(j)|}
\end{equation}
The probability that users who experienced $j$ never experienced $i$ can be modeled as the complementary probability of Equation (\ref{eqn:sym_prob}) w.r.t. $\mathcal{U}(i)$ over $|\mathcal{U}(i) \cup \mathcal{U}(j)|$.
This probability, we name \textit{Jaccard Asymmetric Dissimilarity}(JAD) can be formulated as follows:
\begin{equation}\label{eqn:jad}
JAD(i,j)=\frac{|\mathcal{U}(j)|-|\mathcal{U}(i) \cap \mathcal{U}(j)|}{|\mathcal{U}(i) \cup \mathcal{U}(j)|}
\end{equation}
Again, Equation (\ref{eqn:jad}) can be seen in terms of probability as 
\[
p^{JAD}(i,j) = p(\langle {u},r_{{u}j} \rangle \in \mathcal{U}(j)) -	p^{JS}(i,j)
\]
We now propose to modify the original similarity injecting the former negative correction weighted with a parameter $\lambda$ that can be easily customized. The overall similarity,\textit{ Additive Adjusted Jaccard (AAJ)}, is then formulated as:
\begin{equation}\label{eqn:aaj}
AAJ(i,j)= JS(i,j) - \lambda \cdot JAD(i,j)
\end{equation}
where $\lambda$ is a parameter that could depend on many factors such as the number of users and items in the dataset and the intensity of interactions between them. 

Recalling that the overall formula should represent a degree of similarity between the two different items we defined the Multiplicative Adjusted Jaccard as the product of Jaccard similarity with the inverse of Jaccard Asymmetric Dissimilarity (IJAD). In other words, we use $\frac{1}{p^{JAD}(i,j)}$ as a corrective factor for $p^{JS}(i,j)$.
We then define the \textit{Multiplicative Adjusted Jaccard (MAJ)} as:
\begin{equation}\label{eqn:maj}
MAJ(i,j) =\frac{JS(i,j)}{JAD(i,j)} = JS(i,j) \cdot IJAD(i,j)
\end{equation}
In the multiplicative variant, in order to avoid division by zero the minimum value for JAD is set to $\frac{1}{|\mathcal{U}(i) \cup \mathcal{U}(j)|}$.
As mentioned in Section \ref{sec:motivation} a symmetric variant of Jaccard coefficient (named \textit{Jaccard Symmetric Dissimilarity (JSD)}) can be used composing Equation (\ref{eqn:jad}) with the probability that a user $u$ experienced the item $i$ but never experienced the item $j$:
\begin{eqnarray*}
JSD(i,j) &= &\frac{\left(|\mathcal{U}(j)|-|\mathcal{U}(i) \cap \mathcal{U}(j)|\right)+\left(|\mathcal{U}(i)|-|\mathcal{U}(i) \cap \mathcal{U}(j)|\right)}{|\mathcal{U}(i) \cup \mathcal{U}(j)|}\\
& = & \frac{|\mathcal{U}(i)| + |\mathcal{U}(j)| -2\cdot|\mathcal{U}(i) \cap \mathcal{U}(j)|}{|\mathcal{U}(i) \cup \mathcal{U}(j)|}
\end{eqnarray*}

leading to the corresponding probability
\[
p^{JSD}(i,j) = p^{JAD}(i,j) + p^{JAD}(j,i)
\]
Thus the \textit{Symmetric Additive Adjusted Jaccard (S-AAJ)} and \textit{Symmetric Multiplicative Adjusted Jaccard (S-MAJ)} can be defined as follows:
\[
\text{\textit{S-AAJ}}(i,j)= JS(i,j) - \lambda\cdot JSD(i,j)
\]
\[
\text{\textit{S-MAJ}(i,j)}=\frac{JS(i,j)}{JSD(i,j)}
\]

In order to test our idea, we applied all the variants previously introduced for Jaccard similarity to two more popular similarity measures: \textit{Asymmetric Jaccard Similarity (AJS)} and \textit{S\o rensen coefficient (SOR)}. All the derived variants are represented, respectively, in Table \ref{tbl:ajs_variants} and Table \ref{tbl:sor_variants}.
\begin{table}[ht]
\small
	\centering
	\caption{Asymmetric Jaccard considered variants.}
	\begin{tabular}{|c|c|c|}
		\hline
		Short name & Extended &  Formula \\
		\hline
		&&\\
		AJS(i,j) &  Asymm. Jacc. Similarity & $\frac{|\mathcal{U}(i) \cap \mathcal{U}(j)|}{|\mathcal{U}(i)|}$\\
		&&\\
		AJD(i,j) & Asymm. Jacc. Dissimilarity &  $\frac{|\mathcal{U}(j)|-|\mathcal{U}(i) \cap \mathcal{U}(j)|}{|\mathcal{U}(i)|}$ \\ 
		&&\\
		AAAJ(i,j) & Additive Adjusted As. Jacc. &  $AJS(i,j) - \lambda \cdot AJD(i,j)$ \\
		MAAJ(i,j) &  Multiplicative Adjus. As. Jacc. & $AJS(i,j) \cdot IJAD(i,j)$\\
		S-AAAJ(i,j) & Symmetric AAAJ &  $AJS(i,j) - \lambda \cdot AJD(i,j)$ \\ 
		S-MAAJ(i,j) & Symmetric MAAJ &  $AJS(i,j) \cdot IJSD(i,j)$ \\
		\hline
	\end{tabular}
	\label{tbl:ajs_variants}
\normalsize
\end{table}
\begin{table}[ht]
	\small
	\centering
	\caption{S\o rensen similarity considered variants.}
	\begin{tabular}{|c|c|c|}
		\hline
		Short name & Extended &  Formula \\
		\hline
		&&\\
		SOR(i,j) &  S\o rensen Similarity & $\frac{|\mathcal{U}(i) \cap \mathcal{U}(j)|}{|\mathcal{U}(i)| + |\mathcal{U}(j)|}$\\
		&&\\
		ASD(i,j) & Asymm. S\o r. Dissimilarity &  $\frac{|\mathcal{U}(j)|-|\mathcal{U}(i) \cap \mathcal{U}(j)|}{|\mathcal{U}(i)| + |\mathcal{U}(j)|}$ \\ 
		&&\\
		AAS(i,j) & Additive Adjusted As. S\o r. &  $SOR(i,j) - \lambda ASD(i,j)$ \\
		MAS(i,j) &  Multiplicative Ad. As. S\o r. & $SOR(i,j) \cdot ASD(i,j)$\\
		&&\\
		S-AAS(i,j) & Symmetric AAS &  $SOR(i,j) - \lambda \frac{|\mathcal{U}(i)| + |\mathcal{U}(j)| -2\cdot|\mathcal{U}(i) \cap \mathcal{U}(j)|}{|\mathcal{U}(i)| + |\mathcal{U}(j)|}$ \\ 
		&&\\
		S-MAS(i,j) & Symmetric MAS &  $SOR(i,j) \cdot IJSD(i,j)$ \\
		\hline
	\end{tabular}
	\label{tbl:sor_variants}
	\normalsize
\end{table}


All the above metrics have been introduced having in mind a item-kNN approach but, without loss of generality, they can be applied to user-kNN model as well.

\section{Experimental Evaluation}\label{sec:experiments}
\noindent\textbf{Datasets. }
We evaluated the effectiveness of our approach on three datasets belonging to different domains (Music, Books, and Movies), shown in Table \ref{tbl:datastats}.
The \lastfm dataset \cite{Cantador:RecSys2011} corresponds to transactions with Last.fm online music system released in HetRec 2011\footnote{\url{http://ir.ii.uam.es/hetrec2011/}}. It contains social networking, tagging, and music artist listening information from a set of 2K users.
\library represents books ratings collected in the LibraryThing community website. It contains social networking, tagging, and rating information on a [1..10] scale.
\yahoo (Yahoo! Webscope dataset ydata-ymovies-user-movie-ratings-content-v1\_0)\footnote{http://research.yahoo.com/Academic\_Relations} contains movies ratings generated by Yahoo! Movies up to November 2003. It provides content, demographic, ratings information, and mappings to \movielens and \texttt{EachMovie} datasets.
\begin{table}[ht!]
	\caption{Datasets statistics.}
	\begin{tabular}{l r r r r }
		\hline
		Dataset & \#Users &  \#Items &  \#Transactions  & Sparsity\\
		\hline
		Yahoo! Movies & 7642 &  11,916 & 221,367 &  99.76\%\\ 
		LibraryThing & 7279 &  37,232 & 2,056,487 &  99.24\%\\
		Last FM & 1850 &  11,247 & 59,071 &  99.72\%\\ 
		\hline
	\end{tabular}
	\label{tbl:datastats}
	\begin{flushleft}
		\scriptsize
	Columns corresponding to \#Users, \#Items and \#Transactions show the number of users, number of items and number of transactions, respectively, in each dataset. The last column shows the sparsity of the dataset.
\end{flushleft}
\end{table}\vspace{-0.5cm}

\setlength\tabcolsep{4pt} 
\begin{table*}[ht]
	\caption{Comparison in terms of Precision and Aggregate Diversity for User-kNN scheme}\label{tbl:prec_userknn}
	\centering
	\scriptsize
	\renewcommand{\arraystretch}{1.2}
	\vspace{-0.1cm}
	\begin{tabular}{ |l|r r r r r r r|r|r r r r r r r|r r r r r r r|}
		\hline
		 \multicolumn{23}{|c|}{\textbf{Precision - P@10}}\\
		\hline
		& \multicolumn{1}{c|}{\textbf{JS}} & \multicolumn{2}{c|}{\textbf{AAJ}} & \multicolumn{1}{c|}{\textbf{MAJ}} & \multicolumn{2}{c|}{\textbf{S-AAJ}} & \multicolumn{1}{c|}{\textbf{S-MAJ}} & \multicolumn{1}{c|}{\textbf{ASOR}} & \multicolumn{1}{c|}{\textbf{SOR}} & \multicolumn{2}{c|}{\textbf{AAS}} & \multicolumn{1}{c|}{\textbf{MAS}} & \multicolumn{2}{c|}{\textbf{S-AAS}} & \multicolumn{1}{c|}{\textbf{S-MAS}} & \multicolumn{1}{c|}{\textbf{AJS}} & \multicolumn{2}{c|}{\textbf{AAAJ}} & \multicolumn{1}{c|}{\textbf{MAAJ}} & \multicolumn{2}{c|}{\textbf{S-AAAJ}} & \multicolumn{1}{c|}{\textbf{S-MAAJ}} \\ 
		\hline
		{\textbf{Datasets}} & \textbf{P@10} & \textbf{$\lambda$} & \textbf{P@10} & \textbf{P@10} & \textbf{$\lambda$} & \textbf{P@10} & \textbf{P@10} & \textbf{P@10} & \textbf{P@10} & \textbf{$\lambda$} & \textbf{P@10} & \textbf{P@10} & \textbf{$\lambda$} & \textbf{P@10} & \textbf{P@10} & \textbf{P@10} & \textbf{$\lambda$} & \textbf{P@10} & \textbf{P@10} & \textbf{$\lambda$} & \textbf{P@10} & \textbf{P@10}\\ 
		\hline
		LibraryThing & 0.0363 & 0.2 & 0.0582 & \textbf{0.0627} & 0.4 & 0.0364 & 0.0010 & 0.0394 & 0.0363 & 0.2 & 0.0586 & \textbf{0.0629} & 0.2 & 0.0405 & 0.0364 & 0.0364 & 0.2 & 0.0558 & \textbf{0.0603} & 0.2 & 0.0057 & 0.0375 \\ \hline
		Yahoo & 0.0437 & 0.4 & 0.0561 & \textbf{0.0676} & 0.2 & 0.0403 & 0.0433 & 0.0442 & 0.0438 & 0.4 & 0.0573 & \textbf{0.0687} & 0.2 & 0.0568 & 0.0435 & 0.0607 & 0.6 & 0.0529 & \textbf{0.0664} & 0.2 & 0.0378 & 0.0529 \\ \hline
		Last FM & 0.0164 & 0.4 & 0.0242 & \textbf{0.0257} & 0.2 & 0.0037 & 0.0160 & 0.0215 & 0.0166 & 0.4 & 0.0241 & 0.0253 & 0.2 & \textbf{0.0275} & 0.0164 & 0.0248 & 0.2 & 0.0228 & \textbf{0.0323} & 0.2 & 0.0037 & 0.0242 \\ \hline

%

		\hline\hline
		 \multicolumn{23}{|c|}{\textbf{Aggregate Diversity - D@10}}\\
		\hline

		& \multicolumn{1}{c|}{\textbf{JS}} & \multicolumn{2}{c|}{\textbf{AAJ}} & \multicolumn{1}{c|}{\textbf{MAJ}} & \multicolumn{2}{c|}{\textbf{S-AAJ}} & \multicolumn{1}{c|}{\textbf{S-MAJ}} & \multicolumn{1}{c|}{\textbf{ASOR}} & \multicolumn{1}{c|}{\textbf{SOR}} & \multicolumn{2}{c|}{\textbf{AAS}} & \multicolumn{1}{c|}{\textbf{MAS}} & \multicolumn{2}{c|}{\textbf{S-AAS}} & \multicolumn{1}{c|}{\textbf{S-MAS}} & \multicolumn{1}{c|}{\textbf{AJS}} & \multicolumn{2}{c|}{\textbf{AAAJ}} & \multicolumn{1}{c|}{\textbf{MAAJ}} & \multicolumn{2}{c|}{\textbf{S-AAAJ}} & \multicolumn{1}{c|}{\textbf{S-MAAJ}} \\ 
		\hline
		
		{\textbf{Datasets}} & \textbf{D@10} & \textbf{$\lambda$} & \textbf{D@10} & \textbf{D@10} & \textbf{$\lambda$} & \textbf{D@10} & \textbf{D@10} & \textbf{D@10} & \textbf{D@10} & \textbf{$\lambda$} & \textbf{D@10} & \textbf{D@10} & \textbf{$\lambda$} & \textbf{D@10} & \textbf{D@10} & \textbf{D@10} & \textbf{$\lambda$} & \textbf{D@10} & \textbf{D@10} & \textbf{$\lambda$} & \textbf{D@10} & \textbf{D@10}\\ 
		\hline
		LibraryThing & 2136 & 0.2 & \textbf{7367} & 2406 & 0.4 & 2246 & 873 & 2819 & 2083 & 0.2 & \textbf{7330} & 2292 & 0.2 & 5528 & 2171 & 1299 & 0.2 & \textbf{7501} & 2060 & 0.2 & 3331 & 1486 \\ \hline
		Yahoo & 734 & 0.4 & 2070 & 835 & 0.2 & \textbf{2187} & 825 & 936 & 695 & 0.4 & \textbf{2048} & 786 & 0.2 & 1114 & 784 & 451 & 0.6 & 1773 & 717 & 0.2 & \textbf{2175} & 587 \\ \hline
		Last FM & 1449 & 0.4 & 1345 & \textbf{1654} & 0.2 & 1203 & 1460 & 1433 & 1324 & 0.4 & 1324 & \textbf{1625} & 0.2 & 1410 & 1420 & 774 & 0.2 & \textbf{1665} & 1077 & 0.2 & 1207 & 941 \\ \hline
		
	\end{tabular}
	\vspace{-0.5cm}
\end{table*}

\setlength\tabcolsep{4pt} 
\begin{table*}[ht]
	\caption{Comparison in terms of Precision and Aggregate Diversity for Item-kNN scheme}\label{tbl:prec_itemknn}
	\centering
	\scriptsize
	\renewcommand{\arraystretch}{1.2}
	\vspace{-0.1cm}
	\begin{tabular}{ |l|r r r r r r r|r|r r r r r r r|r r r r r r r|}
		\hline
		\multicolumn{23}{|c|}{\textbf{Precision - P@10}}\\
		\hline
		& \multicolumn{1}{c|}{\textbf{JS}} & \multicolumn{2}{c|}{\textbf{AAJ}} & \multicolumn{1}{c|}{\textbf{MAJ}} & \multicolumn{2}{c|}{\textbf{S-AAJ}} & \multicolumn{1}{c|}{\textbf{S-MAJ}} & \multicolumn{1}{c|}{\textbf{ASOR}} & \multicolumn{1}{c|}{\textbf{SOR}} & \multicolumn{2}{c|}{\textbf{AAS}} & \multicolumn{1}{c|}{\textbf{MAS}} & \multicolumn{2}{c|}{\textbf{S-AAS}} & \multicolumn{1}{c|}{\textbf{S-MAS}} & \multicolumn{1}{c|}{\textbf{AJS}} & \multicolumn{2}{c|}{\textbf{AAAJ}} & \multicolumn{1}{c|}{\textbf{MAAJ}} & \multicolumn{2}{c|}{\textbf{S-AAAJ}} & \multicolumn{1}{c|}{\textbf{S-MAAJ}} \\ 
		\hline
		
		{\textbf{Datasets}} & \textbf{P@10} & \textbf{$\lambda$} & \textbf{P@10} & \textbf{P@10} & \textbf{$\lambda$} & \textbf{P@10} & \textbf{P@10} & \textbf{P@10} & \textbf{P@10} & \textbf{$\lambda$} & \textbf{P@10} & \textbf{P@10} & \textbf{$\lambda$} & \textbf{P@10} & \textbf{P@10} & \textbf{P@10} & \textbf{$\lambda$} & \textbf{P@10} & \textbf{P@10} & \textbf{$\lambda$} & \textbf{P@10} & \textbf{P@10}\\ 
		\hline
		LibraryThing & 0.0869 & 0.2 & \textbf{0.0949} & 0.0451 & 0.4 & 0.0822 & 0.0944 & \textbf{0.1019} & 0.0815 & 0.2 & \textbf{0.0914} & 0.0374 & 0.4 & 0.0826 & 0.0901 & 0.0598 & 0.2 & \textbf{0.1011} & 0.0303 & 0.4 & 0.0830 & 0.0792 \\ \hline
		Yahoo & 0.0331 & 0.2 & 0.0510 & 0.0016 & 0.2 & \textbf{0.0535} & 0.0373 & 0.0447 & 0.0297 & 0.2 & 0.0504 & 0.0014 & 0.2 & \textbf{0.0531} & 0.0352 & 0.0046 & 0.4 & 0.0509 & 0.0012 & 0.2 & \textbf{0.0527} & 0.0083 \\ \hline
		Last FM & 0.0141 & 0.2 & \textbf{0.0248} & 0.0036 & 0.2 & 0.0230 & 0.0158 & 0.0127 & 0.0124 & 0.4 & 0.0195 & 0.0031 & 0.2 & \textbf{0.0230} & 0.0155 & 0.0036 & 0.2 & \textbf{0.0237} & 0.0032 & 0.2 & 0.0229 & 0.0068 \\ \hline
%
%
%
%

	\hline\hline
	\multicolumn{23}{|c|}{\textbf{Aggregate Diversity - D@10}}\\
	\hline
		& \multicolumn{1}{c|}{\textbf{JS}} & \multicolumn{2}{c|}{\textbf{AAJ}} & \multicolumn{1}{c|}{\textbf{MAJ}} & \multicolumn{2}{c|}{\textbf{S-AAJ}} & \multicolumn{1}{c|}{\textbf{S-MAJ}} & \multicolumn{1}{c|}{\textbf{ASOR}} & \multicolumn{1}{c|}{\textbf{SOR}} & \multicolumn{2}{c|}{\textbf{AAS}} & \multicolumn{1}{c|}{\textbf{MAS}} & \multicolumn{2}{c|}{\textbf{S-AAS}} & \multicolumn{1}{c|}{\textbf{S-MAS}} & \multicolumn{1}{c|}{\textbf{AJS}} & \multicolumn{2}{c|}{\textbf{AAAJ}} & \multicolumn{1}{c|}{\textbf{MAAJ}} & \multicolumn{2}{c|}{\textbf{S-AAAJ}} & \multicolumn{1}{c|}{\textbf{S-MAAJ}} \\ 
		\hline
		
			{\textbf{Datasets}} & \textbf{D@10} & \textbf{$\lambda$} & \textbf{D@10} & \textbf{D@10} & \textbf{$\lambda$} & \textbf{D@10} & \textbf{D@10} & \textbf{D@10} & \textbf{D@10} & \textbf{$\lambda$} & \textbf{D@10} & \textbf{D@10} & \textbf{$\lambda$} & \textbf{D@10} & \textbf{D@10} & \textbf{D@10} & \textbf{$\lambda$} & \textbf{D@10} & \textbf{D@10} & \textbf{$\lambda$} & \textbf{D@10} & \textbf{D@10}\\ 
		\hline
		LibraryThing & 9745 & 0.2 & 12004 & \textbf{17510} & 0.4 & 12304 & 10249 & 10399 & 9556 & 0.2 & 11727 & \textbf{16557} & 0.4 & 13306 & 9998 & \textbf{17361} & 0.2 & 11221 & 11132 & 0.4 & 12315 & 14819 \\ \hline
		Yahoo & 3103 & 0.2 & 3718 & 3463 & 0.2 & \textbf{4315} & 2541 & 3346 & 3184 & 0.2 & \textbf{3935} & 3397 & 0.2 & 3541 & 2475 & 2948 & 0.4 & \textbf{4092} & 2334 & 0.2 & 3756 & 2295 \\ \hline
		Last FM & 3508 & 0.2 & \textbf{3538} & 3486 & 0.2 & 3462 & 2911 & \textbf{3779} & 3265 & 0.4 & 305 & 3175 & 0.2 & 3509 & 2887 & 2992 & 0.2 & \textbf{3935} & 2644 & 0.2 & 3692 & 3457 \\ \hline
	\end{tabular}
\end{table*}

\noindent\textbf{Evaluation Protocol and Experimental Setting. }
We employed the \textit{all unrated items} \cite{Steck13} evaluation protocol to evaluate the proposed method. In \textit{all unrated items} the recommendation list is computed from a candidate list given by the cartesian product between users and items minus the items each user experimented in the training set. We performed a temporal 80-20 hold-out split retaining the last 20\% of ratings as test set when temporal information is available.
We measured the performance by computing Precision@N ($Prec@N$) for \TopN recommendation lists as accuracy metric.  Precision has been computed on a per-user basis and then results have been averaged. As Precision needs relevant items to be computed, we set the relevance threshold to 8 over 10 for \library and \yahoo, and to 0 for \lastfm since in the latter no ratings are provided but the number of user-item transactions. We measured Diversity through \textit{catalog coverage} (aggregate diversity in \TopN list). The \textit{catalog coverage}, also called diversity-in-top-N ($D@N$) \cite{AdomaviciusK12}, is measured by computing the overall number of different items recommended within all recommendation lists. It represents the propensity of a system to recommend always the same items.\\
\noindent\textbf{Baselines. }
We compared our approaches with in both User-kNN and Item-kNN settings. The former finds the k-nearest user neighbors based on a similarity function and then exploits them to predict a score for each user-item pair. The latter is the item-based version of the k-nearest neighbors algorithm that uses the k-nearest items to make the predictions.
For both the algorithms $k=80$ has been set. However, we are not interested in the algorithm itself but on the similarity measures that are used to compute neighbors and predictions. As baseline to compare with, we used both symmetric and asymmetric measures, namely, \textbf{\textit{Jaccard (JS)}} and \textbf{\textit{Sorensen (SOR)}} (for symmetric measures) and  \textbf{\textit{asymmetric Jaccard (AJS)}} and \textbf{\textit{asymmetric Jaccard weighted with the Sorensen Index (ASOR)}} \cite{katarya2016effectivecollaborative}  (for asymmetric measures).
For all the similarities that make use of $\lambda$  we evaluated them varying $\lambda$ in {0.2,0.4,0.6,0.8}. Only best results have been shown in Tables \ref{tbl:prec_userknn} and \ref{tbl:prec_itemknn}\footnote{The complete results are publicly available at \url{https://github.com/sisinflab/The-importance-of-being-dissimilar-in-Recommendation}.}. The best values in terms of Precision and Aggregate Diversity are highlighted in bold. We computed significance tests for precision highlighted results and we found them statistically significant at the 0.05 level w.r.t. their respective baselines.

\noindent\textbf{Performance of the proposed methods. }
The results in Table \ref{tbl:prec_userknn} show that our approach always outperforms baseline variants in User-kNN scheme. In details \textbf{additive} asymmetric similarity and \textbf{multiplicative} asymmetric similarity significantly perform better than JS, SOR and ASOR for all three dataset. Among these two variants of similarity, the multiplicative variant is absolutely the best-performing one. Quite interestingly, modifying AJS, which is asymmetric in its inner nature with our asymmentric dissimilarity factor leads to an improvement irrespective of the considered dataset. 
It is worth noticing that, other than the accuracy improvements, aggregate diversity also increases due to the dissimilarity injection. In details, the asymmetric additive variant achieves the best results and triples catalog coverage values for \library and \yahoo.

Table \ref{tbl:prec_itemknn} shows Precision and Catalog Coverage results for an item-based scheme. These results are very interesting for many reasons. First of all, it is clear that, same similarities can lead to very different results depending on the scheme adopted. In particular asymmetric Jaccard (AJS) performs very badly for item-kNN algorithm. Under the dissimilarities perspective, it happens exactly the same and multiplicative approach performs bad. Quite surprisingly the additive version is able to always outperform the base variants. This suggests that adopting an additive strategy for item-kNN may lead to better results. This is probably due to the wide number of items pairs without any common user. Focusing on additive symmetric and asymmetric similarities we can note that aggregate diversity results reflect the same improvements observed in accuracy results. The only case that appear to behave differently is AAAJ that registered a catalog coverage lower than AJS. Actually, this happens as we considered the best performing $\lambda$ in terms of precision. In case of $\lambda \in \{0.4,0.6\}$ we obtain aggregate diversity values of 16,205 and 18,071, respectively, with precision values constantly higher than AJS (0.09374 and 0.08820).

\section{Conclusion and Future Work}\label{sec:conclusion}
In this work we propose a method to improve performance of neighborhood based models, by capturing subtle interactions between users and items, which cannot be appreciated using a traditional similarity measure. We defined a dissimilarity measure, that can be used combined with traditional user-based and item-based schemes. The proposed approach takes into account the single asymmetric components, leading to more fine improvements in terms of precision and aggregate diversity. We performed a comparative experimental evaluation using three well-known datasets, varying the tuning parameter $\lambda$. Experiments show that our approach outperforms competing algorithms, denoting the usefulness of incorporating symmetric and asymmetric dissimilarity in neighborhood-based models.
We are currently working on an extension of our idea that takes into account also user ratings and not just set-based measures. As further extension we are also interested in making the approach even more personalized by weighting dissimilarity with user-centered lambdas. 


\bibliographystyle{ACM-Reference-Format}
\bibliography{bibliographywa}

\end{document}